\documentclass{elsart}
\usepackage{amssymb}
\usepackage{amsmath}
\usepackage[dvips]{graphicx}

\journal{:\bf \ \ European Journal of Mechanics, B/Fluids,
\textbf{15}, 4, pp. 545-568
(1996).\qquad\qquad\qquad\qquad\qquad\qquad\qquad\qquad\qquad\qquad\qquad
}
\input{tcilatex}
\begin{document}

\begin{frontmatter}

\title{Nucleation of spherical
shell-like interfaces by second gradient theory: numerical
simulations}

\author{Francesco DELL'ISOLA}
\address {
Universit\`a  di Roma La Sapienza, Dipartimento di
Ingegneria  Strutturale e Geotecnica, Via Eudossiana, 18 - 00184
Roma, Italy} \ead{francesco.dellisola@uniroma1.it}
\author{Henri GOUIN}

\address {
   Universit\'e d'Aix-Marseille \& C.N.R.S.  U.M.R. 6181\\ Case 322,
Av. Escadrille
 Normandie-Niemen, 13397 Marseille Cedex 20, France}
\ead  {henri.gouin@univ-cezanne.fr}

\author{Giacomo ROTOLI}
\address {Universit\`a
dell'Aquila, Dipartimento di Energetica, Facolt\`a di Ingegneria, \\ Roio Poggio 67040, L'Aquila, Italy}
\ead{rotoli@ing.univaq.it}

\begin{abstract}
 The theory of second gradient fluids (which are able to exert shear
stresses also in equilibrium conditions) allows us: (i) to describe
both the thermodynamical and the mechanical behavior of systems in
which an interface is present; (ii) to express the surface tension
and the radius of microscopic bubbles in terms of a functional of
the chemical potential; (iii) to predict the existence of a (minimal)
nucleation radius for bubbles. Moreover, the above theory supplies a
3D-continuum model which is endowed with sufficient structure to
allow, using the procedure developed by Dell'Isola \& Kosi\'{n}ski, the
construction of a 2D-shell-like continuum representing a consistent
approximate 2D-model for the interface between phases. In this paper
we use numerical simulations in the framework of second gradient
theory to obtain explicit relationships for the surface quantities
typical of 2D-models. In particular, for some of the most general
two-parameter equations of state, it is possible to obtain the
curves describing the relationship between the surface tension, the
thickness, the surface mass density and the radius of the spherical
interfaces between fluid phases of the same substance. These results
allow us to predict the (minimal) nucleation radii for this class of
equations of state.
\end{abstract}


\end{frontmatter}

\section{Introduction}

In 1949 Tolman, using the thermodynamical approach due to Gibbs, established
a relationship between surface tension and radius of bubbles in equilibrium
with their liquid phase \cite{Tolman1}. It fails when the bubble has a
radius close to the nucleation value \cite{Lamer}: indeed by following the
ideas of Gibbs \cite{Gibbs} it is impossible to prove the existence of a
'minimal' nucleation radius, i.e. the minimal radius of a bubble or droplet
necessary for nucleation in the other phase. Cahn and Hilliard \cite{Cahn}
proposed more sophisticated models in which the nucleation phenomena are
accounted for. They introduce a 3D-description of the interfacial region,
which is characterized as the region in which high mass density gradients
occur, so that a supplementary term is added to the classical free energy
expression. This term is in general non-local, i.e. it depends on all
gradients of mass density \cite{Gouin1}. It can be shown however that in
many cases the first gradient\footnote{%
In French literature this theory is called 'second' gradient rather than
'first' gradient after Germain \cite{Germain}.} of mass density is
sufficient to give an excellent approximation to the exact solution \cite%
{Falls}. This approach was developed in the literature mainly using the
methods of statistical mechanics \cite{Davis,de Gennes,Evans,van Kampen} but
also in the framework of classical field theories \cite%
{Casal3,Germain,Serrin2} by means of the so-called second gradient theory.
The second gradient fluids (or capillary fluids as named by Casal \& Gouin
\cite{Casal3}) are fluids which are able to exert shear stresses in
equilibrium conditions and contact forces concentrated on lines \cite%
{Seppecher}. The theory also takes into account the mechanical aspects of
the phenomena considered in the theories of Laplace, Gibbs and Tolman.
Indeed it is easy to prove, for second gradient fluids, that the pressure
field in equilibrium conditions may be non-constant. On the contrary it is
an essential feature of the classical Laplace theory that the pressure field
is spatially uniform. Moreover in the framework of second gradient theory it
is possible to deduce the PDE governing the (equilibrium) mass density
distribution from the force balance law. Dell`Isola et al. \cite{Isola1}
show the consequences of the distinction (first noticed by Rocard \cite%
{Rocard}), between thermodynamic pressure (that deriving from free energy)
and the mechanical pressure (i.e. the trace of the capillary stress tensor).
Indeed in the quoted papers it is shown that a definition of the surface
tension and of the radius of a bubble is possible also in the 3D-approach by
using the second gradient theory and by means of the concept of the
'equivalent Laplace bubble'. In this way it is possible to obtain
theoretical predictions for minimal nucleation radii and dependence between
the surface tension and the bubble radius; moreover it is possible to use
the 'equivalent Laplace bubble' to study the stability of the critical
nuclei \cite{Gouin3}. Unfortunately in the case of spherical bubbles the ODE
determining mass density profile cannot be easily studied, so that it is
necessary to use numerical simulations in order to be able to make a
theoretical prediction. We explicitly remark that in the present paper a
'shell-like' model for the interface is used, following the ideas developed
in \cite{Isola2}. In this approach there is no use of 'Gibbs excesses'
quantities (for a detailed treatment of these concepts see for instance \cite%
{Chattoraj}), so our paper substantially differs from \cite{Falls} in which
an analogue analysis for drops is developed. The main result of this paper
is a new relation between surface tension and the radius of bubbles which
holds also for microscopic radii and which allows the determination of the
minimal nucleation radius. The paper is organized as follows: in Section 2
we collect the main results due to Gibbs-Tolman, Korteweg, Dunn-Serrin,
Gurtin and Cahn-Hilliard and then we compare the results with experiments;
in Section 3 we state the theoretical framework of second gradient theory
and its application to bubbles; in Section 4 we study the equation for mass
density profiles for some of the most general two-parameter equations of
state and solve it numerically; in Section 5 comparisons with experiments,
together with an explicit prediction about 'minimal nucleation radii' of
some substances, are carried out; finally some conclusions are collected in
Section 6.

\section{Model of interface between fluid phases: the problem of nucleation
of microscopic bubbles}

\subsection{Nucleation of microscopic bubbles}

In \cite{Isola3,Isola4} it is shown that Tolman formula \cite{Tolman1} is an
universal property of 2D-Gibbs models for the interface between phases. In
fact the Tolman formula is a consequence of the Laplace formula and some
simple assumptions (whose origin will be found in the works of Gibbs \cite%
{Gibbs}) about the thermodynamics of 2D-continua. These assumptions are
basically: (i) the possibility of defining surface quantities such as
surface mass density, surface free energy, etc.; (ii) the validity of
standard thermodynamical relationship between these quantities. The Tolman
formula predicts that for microscopic bubbles or droplets, the equilibrium
surface tension will decrease exponentially as a function of the equilibrium
radius. However very soon after its publication in 1949 Tolman's formula was
criticized by Lamer and Pound \cite{Lamer} who, founding their analysis on
nucleation data for very small droplets, found no significant decrease of
surface tension with the radius. The experimental failure of Tolman theory
has substantially remained unjustified theoretically until now \cite%
{Fisher,Kumar}. In our opinion two facts are relevant to this problem: (i)
the failure of any 'classical' theory is a consequence of idealizing the
interface as a region without thickness; surface quantities would be
correctly defined stating a 3D-theory of interfaces and then integrating
over the interface thickness in order to define surface quantities; (ii) the
failure of Tolman formula for droplets/bubbles of very small radii cannot be
surprising because in the Tolman theory no prediction of the 'minimal'
critical nucleation radius is made, so the Tolman-Gibbs theory cannot
explain an essential physical phenomenon. Concerning the point (i) many
authors have tried to define surface quantities as an integral over the
interface: the first very crude attempt is due to Tolman himself \cite%
{Tolman2}; many authors are able to define surface tension in term of mass
density profile, or its gradient, by following either the ideas stemming
from \cite{Cahn} (cf. also \cite{Davis,Evans,Rocard}), though these
results are applicable only for large bubbles/droplets, or making use of the
Gibbs concept of a dividing surface \cite{Falls}; a more rigorous approach,
using a shell-like approach, is followed in \cite{Isola2}. Point (ii) is
closely connected to point (i): the description of homogeneous nucleation
phenomena is possible only in a 3D-theory of the interface because we have
to `form` the interface with the critical nucleus starting from a single
phase. The idealized model of the interface as a simple two-dimensional
continuum can obviously fail if the radius of the microscopic bubble (or
droplet) and the interface thickness are of the same order of magnitude. It
will be shown below that for typical nucleation problem a 2D-theory is not
suitable as, in that case, the vapor bubble surrounded by its liquid phase
is constituted mainly or exclusively by the 'interfacial region'. As it is,
however, still important to attribute a radius and an energy to
microscopical bubble, we do it by means of an \textit{equivalent} model of
Laplace-Gibbs type. In this way it becomes possible to interpret the
experimental evidence.

\subsection{Cahn-Hilliard-Korteweg capillary fluids}

The Cahn-Hilliard theory was developed just to account for the nucleation or
more generally to study the behavior of strongly inhomogeneous fluids. In
this theory a term depending on the gradient of the mass density is added to
the classical expression for the free energy valid for homogeneous continua.
Usually, the additional term is assumed to depend only on the gradient of
density \cite{Cahn, Serrin2}. Van Kampen has obtained the same results using
the ideas and methods of statistical mechanics applied to a classical van
der Waals gas \cite{van Kampen}. We remark that in general, gradient
theories are in many cases an excellent approximation of non-local theory,
in which the additional term in the free energy depends on all the gradients
of the mass density and this can be derived only by solving an integral
equation \cite{Falls,van Kampen}. On the other hand we note that Korteweg,
using the ideas of continuum mechanics, proposed some constitutive laws for
the stress tensor which allowed him to prove the Laplace law \cite{Korteweg}%
. However, as proven by Gurtin, classical Cauchy materials cannot show
behavior of the Korteweg type, because of the second principle of
thermodynamics \cite{Gurtin}. Dunn and Serrin \cite{Dunn} proposed to change
the expression for the power in Cauchy materials (adding the so-called
"interstitial work") in order to avoid the contradiction shown in \cite%
{Gurtin}. Another possibility could be to change the fundamental assumptions
about contact forces (Cauchy postulate) characterizing Cauchy materials.
When interpreted in the language of the classical continuum mechanics, the
above assumptions about free energy lead to an expression for stress tensor
which differs from that valid for compressible fluids as equilibrium shear
components appear in it. The form of force balance equation valid for second
gradient fluids is found in Germain \cite{Germain}. However controversial,
the introduction of second order fluids seems to be able to model some
interesting problems in a simple way, as underlined by Casal, who named them
"capillary" fluids \cite{Casal1}. We remark that Cahn and Hilliard do not
consider the mechanical aspect of the nucleation phenomenon, which is
accounted for in the Laplace approach. For this reason they only consider
the thermodynamic pressure (i.e. the spherical part of the stress tensor
deriving from the classical expression of free energy) while Casal shows,
for second gradient fluids, the existence of a capillary non-spherical
stress tensor whose trace includes, but does not reduce to, the quoted
thermodynamic pressure \cite{Casal2}. Finally we remark that Cahn and
Hilliard introduced, using their 3D-approach, an expression for surface
tension valid only for plane interfaces, and never investigated the
relationship between surface tension and radius (see also \cite{Davis,Evans}%
).

\subsection{Experimental evidence}

From an experimental point of view the failure of Tolman formula was
verified by Fisher and Israelachvili \cite{Fisher}, though their results
deserve some comments: their fundamental result is the observation that no
significative variation of surface tension was observed down to radii near
to the molecular diameter. However their experimental apparatus cannot
reproduce a process of homogeneous nucleation because at such small radii
the adhesive effects of their 'hyperboloidal' bubble can significantly
change the boundary value problem for the density profile. On the other hand
the 'minimal' nucleation radius predicted from the Cahn and Hilliard theory
or by second gradient theory appears typically larger than the molecular
diameter by a factor of two or three, so reproducing the data typical of
nucleation studies (see Section 5 below). Another problem connected with the
Fisher and Israelachvili experiment is that radii are obtained via the
Kelvin relation that (though surely 'verified down to 4 $nm$' as quoted by
the authors) needs some correction near to the 'minimal' nucleation radius.
We think that it is possible to reinterpret their data by using the \textit{%
equivalent} model described in the next section. More recently Kumar et al
\cite{Kumar} proposed a slight modification of Tolman results that can
account for some of the observed nucleation data: a comparison of second
gradient theory with their values is reported below in Section 5 for some
data; here we note that the Kumar theory cannot predict a 'minimal'
nucleation radius for droplets (the 'critical' radius data in their paper
will be interpreted more correctly as an unstable 'equilibrium' radius \cite%
{van Carey}. The reason is that their theory is essentially a modification
of the 2D-Laplace-Gibbs theory in which the interface has no thickness.

\section{Shell-like interfaces by second gradient theory}

\subsection{Equilibrium equations for a fluid endowed with internal
capillarity}

The second gradient theory, conceptually more straightforward than Laplace's
theory, can be used to construct a theory of capillarity. We aim to study
the equilibrium properties of a system with an interface under fixed
temperature (isothermal conditions). In the following whenever stated
capital letters $P$, $W$, etc. denote classical (per unit volume if the
quantity is extensive) thermodynamic quantities; small Greek letters $%
\epsilon $, $\mu $ denote classical mass-based thermodynamic quantities. For
example, if $\psi $ is the classical mass free energy, the volume free
energy will be $W=\rho \psi $, and the classical chemical potential $\mu $
is given by:
\begin{equation*}
\mu (\rho )=\frac{\partial W}{\partial \rho }.\eqno(3.1)
\end{equation*}%
In the present text the only addition is an internal mass energy $\epsilon $
that is a function of the density $\rho $ as well as $\nabla \rho $. The
internal mass energy characterizes both the compressibility and capillarity
properties of the fluid, independently of the bodies with which it is in
contact. For an isotropic fluid, it is assumed that:
\begin{equation*}
\epsilon =\epsilon (\rho ,\beta _{\rho }),\eqno(3.2)
\end{equation*}%
where $\beta _{\rho }=(\nabla \rho )^{2}$. The equation of equilibrium is
written:
\begin{equation*}
\mathrm{div}\,\mathbf{S}-\rho \nabla \Omega =0,\eqno(3.3)
\end{equation*}%
where $\Omega $ is the extraneous force potential and $\mathbf{S}$ is the
general stress tensor:
\begin{equation*}
\mathbf{S}=-p\,\mathbf{I}-\lambda (\nabla \rho )\otimes (\nabla \rho )^{T},%
\eqno(3.4)
\end{equation*}%
with $\lambda =2\rho \,\epsilon _{,\beta _{\rho }}$ and $p=\rho ^{2}\epsilon
_{,\rho }-\rho \func{div}(\lambda \nabla \rho )$. Following Rocard, this
last expression for $p$ is what we call the \textit{mechanical pressure}
\cite{Rocard}\footnote{%
We remark that in \cite{Falls} only the first term $\rho ^{2}\epsilon
_{,\rho }$, corresponding to the classical thermodynamic pressure $P$,
enters in the Laplace law, cf. Eq. (4.9) of the quoted reference; however in
appendix A the authors agree that the mechanical pressure needs
consideration.}.

The equation of equilibrium (3.3) is obtained in the clearest manner by
using classical methods of the variational calculus, i.e. using the virtual
work Principle due to d'Alembert-Lagrange (cf. Serrin \cite{Serrin1}). It
appears that a single term containing $\lambda $ accounts of the second
gradient effects in the equilibrium equation. The quantity $\lambda $, like $%
\epsilon $, depends on $\rho $ and $\nabla \rho $ (see also the discussion
of the predictions of statistical mechanics on the $\rho $ dependence in
Section 4 below and references cited therein). In a study of surface tension
based on gas kinetic theory \cite{Rocard}, Rocard obtained the same
expression (3.4) for the stress tensor, but with $\lambda $ constant. If $%
\lambda $ is constant, the internal energy is simply written:
\begin{equation*}
\epsilon (\rho ,\beta _{\rho })=\epsilon _{0}(\rho )+\frac{\lambda }{2\rho }%
\,\beta _{\rho },\eqno(3.5)
\end{equation*}%
i.e., the second gradient term $\displaystyle\frac{\lambda }{2\rho }\,\beta
_{\rho }$ is simply added to the energy $\epsilon _{0}(\rho )$ of the
classical compressible fluid. The thermodynamic pressure for the fluid is $%
P=\rho ^{2}\epsilon _{0,\rho }$, which gives for the mechanical pressure $p$%
:
\begin{equation*}
p=P-\lambda \left( \frac{1}{2}\beta _{\rho }+\rho \Delta \rho \right) ,\eqno%
(3.6)
\end{equation*}%
where $\Delta $ is the Laplace operator. This pressure appears in the
boundary conditions \cite{Seppecher}. For $P$, Rocard uses the van der Waals
pressure or two-phase state laws.

In isothermal equilibrium conditions, Eq. (3.3) can also written as:
\begin{equation*}
\mu (\rho )-\lambda \Delta \rho +\Omega =const.\;,\eqno(3.7)
\end{equation*}%
where $\mu $ is given by Eq. (3.1), i.e., is the chemical potential,
relative to the fluid in an homogeneous state with mass density $\rho $ at a
fixed temperature $T$. If $\Omega $ is negligible the equilibrium of a
liquid bubble\footnote{%
The same approach could be followed also in the case of vapor droplets, but
here for sake of brevity we limit ourselves to develop the theory for
bubbles.} surrounded by its liquid phase with density $\rho _{l}$ is
represented by a spherically symmetric density profile satisfying:
\begin{equation*}
\lambda \left( \rho _{,rr}+\frac{2}{r}\rho _{,r}\right) =\mu (\rho )-\mu
(\rho _{l}),\eqno(3.8)
\end{equation*}%
We call Eq. (3.8) the Density Profile Equation (DPE). To the DPE we have to
add boundary conditions: \newline
(i) because of spherical symmetry the derivative of $\rho $ vanishes at the
origin;\newline
(ii) as we have assumed that the bubble is surrounded by an homogeneous
liquid the same condition holds at infinity\footnote{%
If $\rho _{l}^{P}$ is the density of the liquid phase in the equilibrium
state with plane interface and if the constitutive function $\mu $ is
suitably regular, the theory of Fuchs equations implies that:\newline
(iii) for every $\rho _{l}$ in an interval $(\rho _{spi},\rho _{l}^{P})$,
where $\rho _{spi}$ is the spinodal density value, the DPE and conditions
(i)-(ii) determine uniquely an increasing mass density profile $\rho (r)$ and,
in particular, the value $\rho _{v}$ at the origin; we will say that $\rho
(r)$ satisfies the capillary fluid version of Gibbs Phase Rule; \newline
(iv) $\rho (r)$ is twice differentiable at the origin;\newline
v) as $\left( \frac{d\mu }{d\rho }\right) _{\rho _{l}}>0$, $\rho (r)$
converges at least exponentially to $\rho _{l}$ when $r$ tends to infinity.}.%
\newline
Equation (3.8) was formulated by Rocard \cite{Rocard} and by Cahn \&
Hilliard \cite{Cahn} (cf. also Blinowski \cite{Blinowski}, Truskinovsky \cite%
{Truskinovsky} and for a more general mathematical treatment Peletier \&
Serrin \cite{Peletier}).

With $W$, the volume free energy being a primitive of $\mu $, by multiplying
the DPE by $\rho _{r}$ and integrating between zero and $r$ we obtain:
\begin{equation*}
\frac{\lambda }{2}\rho _{,r}^{2}+2\lambda \int_{0}^{r}\frac{\rho _{,r}^{2}}{r%
}dr=W(\rho )-W(\rho _{v})-\mu (\rho _{l})(\rho -\rho _{v}).\eqno(3.10)
\end{equation*}%
Moreover by multiplying the DPE by $r$ and integrating again between zero
and $r$, we obtain:
\begin{equation*}
\lambda r\rho _{r}(r)+\lambda (\rho (r)-\rho _{v})=\int_{0}^{r}[\mu (\rho
)-\mu (\rho _{l})]rdr,
\end{equation*}%
and by taking the limit for $r\,\rightarrow \,\infty $, $r\rho _{r}(r)$
tends to zero and we obtain:
\begin{equation*}
\lambda (\rho _{l}-\rho _{v})=\int_{0}^{\infty }[\mu (\rho )-\mu (\rho
_{l})]rdr.\eqno(3.11)
\end{equation*}%
Such expressions as Eqs. (3.10)-(3.11) can be useful to test the numerical
calculations of the interface profile, especially the second one because it
relates the density jump across the interface directly to a simple integral
involving chemical potential.

\subsection{Nucleation energy of bubbles}

In the theory of Laplace-Gibbs we have for bubbles of radius $R$ and surface
tension $\sigma $:
\begin{equation*}
w=4\pi R^{2}\sigma +\frac{4}{3}\pi R^{3}\big(W(\rho _{v})-W(\rho _{l})-\mu
(\rho _{l})(\rho _{v}-\rho _{l})\big),\eqno(3.12)
\end{equation*}%
with $P=\rho \mu -W$ being the thermodynamic pressure; the isothermal
equilibrium conditions:
\begin{equation*}
\mu (\rho _{l})=\mu (\rho _{v}),
\end{equation*}%
\begin{equation*}
P(\rho _{l})-P(\rho _{v})=-\frac{2\sigma }{R},
\end{equation*}%
are valid only in the context of Laplace theory and transform Eq. (3.12)
into:
\begin{equation*}
w=\frac{4}{3}\;\pi R^{2}\sigma .\eqno(3.13)
\end{equation*}%
One can conclude that the nucleation energy of the bubble is one third of
the creation energy of its interface.\newline
We will now extend this result to the theory of second gradient fluids: in
the theory of second gradient fluids, by following Cahn and Hilliard \cite%
{Cahn}, we have for the nucleation energy of a bubble in a domain $D$ (cf.
also Dell'Isola et al \cite{Isola4}):
\begin{equation*}
w=\int_{D}\left[ W(\rho )-W(\rho _{l})-\mu (\rho _{l})(\rho -\rho _{l})+%
\frac{\lambda }{2}(\nabla \rho )^{2}\right] dv.\eqno(3.14)
\end{equation*}%
By denoting $\psi (\rho )=W(\rho )-W(\rho _{l})-\mu (\rho _{l})(\rho -\rho
_{l})$, by multiplying Eq. (3.10) by $r^{2}$ and by integrating the result
over $[0,+\infty ]$ we get:
\begin{equation*}
\int_{0}^{\infty }\left[ \frac{\lambda }{2}\rho _{,r}^{2}-\psi (\rho )\right]
r^{2}dr+\int_{0}^{\infty }\left[ \psi (\rho _{v})+\int_{0}^{r}\frac{2\lambda
}{r}\rho _{,r}^{2}dr\right] r^{2}dr=0.
\end{equation*}%
By integrating by parts and using Eq. (3.10) again, this equation becomes:
\begin{equation*}
\int_{0}^{\infty }\left[ \frac{\lambda }{6}\rho _{,r}^{2}+\psi (\rho )\right]
r^{2}dr+\left[ \frac{r^{3}}{3}\left( \psi (\rho )-\frac{\lambda }{2}\rho
_{,r}^{2}\right) \right] _{0}^{\infty }=0.
\end{equation*}%
Because of properties of the density profile quoted in Section 3.1, the
second term vanishes, and we get:
\begin{equation*}
w=4\pi \int_{0}^{\infty }\left[ \psi (\rho )+\frac{\lambda }{2}\rho _{,r}^{2}%
\right] r^{2}dr=\frac{4}{3}\pi \int_{0}^{\infty }\lambda \rho
_{,r}^{2}r^{2}dr.\eqno(3.15)
\end{equation*}%
By denoting $\overline{R}^{2}$ the mean value of $r^{2}$ with respect to the
measure $\rho _{,r}^{2}dr$, the above equation reads:
\begin{equation*}
w=\frac{4}{3}\pi \overline{R}^{2}\int_{0}^{\infty }\lambda \rho _{,r}^{2}dr.%
\eqno(3.16)
\end{equation*}%
For a large bubble, i.e., when $\rho _{l}$ tends to the plane interface
value, $\overline{R}$ represents the radius. The surface tension for a plane
interface is $\displaystyle\int_{0}^{\infty }\lambda \rho _{r}^{2}dr\ $ \cite%
{Cahn,Davis,Evans}. Therefore Eq. (3.16) extends Eq. (3.13) to microscopic
bubbles and reduces to it in case of large bubbles.

\subsection{Comparison of Laplace and second gradient theories. Equivalent
Laplace bubbles}

In the second gradient theory the stress tensor in the centre of a spherical
bubble takes the value $\mathbf{S}=-p_{v}\mathbf{I}$ where $p_{v}=P(\rho
_{v})-\lambda \rho _{v}\ \Delta \rho _{|\rho =\rho _{v}}$. The value $\rho
_{l}$ of the mass density in the liquid phase is attained asymptotically
while the stress tensor takes the value $\mathbf{S}=-p_{l}\mathbf{I}$ where $%
p_{l}=P(\rho _{l})$. As the DPE implies $\lambda \rho _{v}\ \Delta \rho
_{\rho =\rho _{v}}=\rho _{v}[\mu (\rho _{v})-\mu (\rho _{l})]$ we have:
\begin{equation*}
p_{v}-p_{l}=W(\rho _{l})-W(\rho _{v})+\mu (\rho _{l})(\rho _{v}-\rho _{l}).%
\eqno(3.17)
\end{equation*}%
Let us notice that this difference is \textit{not equal} to the
corresponding difference of the thermodynamical pressures as, for
microscopic bubbles, $\mu (\rho _{l})$ differs from $\mu (\rho _{v})$. As
the experimental results (see for instance Fisher \& Israelachvili \cite%
{Fisher}) deal with measurements of stresses, then we have to use $%
p_{v}-p_{l}$ instead of $P(\rho _{v})-P(\rho _{l})$ in the comparison
between Laplace theory and second gradient theory. We can now define the surface
tension and the radius of a bubble by identifying the nucleation energies
and the pressure differences computed in both theories as follows:
\begin{equation*}
p_{v}-p_{l}=\frac{2\sigma }{R}\qquad \mathrm{and}\qquad \frac{4}{3}\pi
R^{2}\sigma =\frac{4}{3}\pi \int_{0}^{\infty }\lambda \rho _{,r}^{2}r^{2}dr,
\end{equation*}%
which leads to the relations
\begin{equation*}
R=\left[ 2\lambda \int_{0}^{\infty }\rho _{,r}^{2}r^{2}dr\right] ^{\frac{1}{3%
}}\left[ W(\rho _{l})-W(\rho _{v})-\mu (\rho _{l})(\rho _{l}-\rho _{v})%
\right] ^{-\frac{1}{3}},\eqno(3.18)
\end{equation*}%
and
\begin{equation*}
\sigma =\left[ \frac{\lambda }{4}\int_{0}^{\infty }\rho _{,r}^{2}r^{2}dr%
\right] ^{\frac{1}{3}}\left[ W(\rho _{l})-W(\rho _{v})-\mu (\rho _{l})(\rho
_{l}-\rho _{v})\right] ^{\frac{2}{3}}.\eqno(3.19)
\end{equation*}%
Equations (3.18)-(3.19) define the radius and the surface tension of a
microscopic 'equivalent Laplace bubble' in our 'shell-like' theory. We
remark that they differ from Eq. (4.9) of Falls et al \cite{Falls};
moreover, in Falls et al, $R$ is not specified in terms of the density
profile. Indeed Figures 18 and 19 of the quoted reference are drawn by means
of the 'Gibbs excesses' approach, so that these plots have in abscissa the
Gibbsian 'radius of tension' or 'surface of tension' \cite{Chattoraj}.

\subsection{Thickness and surface mass density of interfaces}

One of the fundamental problems of interfaces is the study of their
thickness. We note that in the Laplace-Gibbs model the interfaces have no
thickness. In the above proposed model, in which $\lambda$ is constant, the
interface thickness is infinite. This is a consequence of the fact that the
limit density in the phases is reached asymptotically. On the other hand the
experiments show that the interfaces generally have a very small thickness,
i.e., typically a few molecular diameters. The thickness takes a macroscopic
dimension when the critical point is approached or the density is almost
uniform (near the so-called spinoidal limit \cite{Cahn}). Therefore a sort
of coherence length for the interface is needed in second gradient theory.
Some natural expressions can be proposed to characterize the thickness of
the interface taking into account only the region in which the strongest
gradient occurs. The most simple of these expressions is the quadratic
variance associated to a measure $\rho_{,r}^2r^2dr$ (we note that the first
order variance is zero). We propose:
\begin{equation*}
\delta^2=\frac{\displaystyle\int_0^{\infty}(r-R)^2\rho_{,r}^2r^2dr} {%
\displaystyle\int_0^{\infty}\rho_{,r}^2r^2dr}. \eqno(3.20)
\end{equation*}
We call $\delta$ the 'equivalent' thickness of the interface. We remark that
there is also a natural definition of the bubble radius:
\begin{equation*}
\displaystyle R_1=\frac{\displaystyle\int_0^{\infty}r\rho_{,r}^2r^2dr}{%
\displaystyle\int_0^{\infty}\rho_{,r}^2r^2dr}. \eqno(3.21)
\end{equation*}
Numerical simulations show that, in all the cases studied below, $R_1$ and $%
R $ are practically the same (differing by less than $1\%$ except very near
to the critical point). In the 2D-interface models the notion of surface
mass density $\rho_{\sigma}$ is introduced. In the treatment proposed in
\cite{Isola2} it is assumed that:
\begin{equation*}
\rho_{\sigma}=\int_{R-\delta/2}^{R+\delta/2}\rho jdr, \eqno(3.22)
\end{equation*}
where $j(r)=1-2H(r-R)+K(r-R)^2$ is the curvature dependent Jacobian
pertaining to the change of variables between surfaces parallel to the
interface in the 2D-models.\newline
In the next Section we study by numerical simulations, for different
two-parameter equations of state, the dependence of the interface thickness
and of the surface mass density as functions of the radius and the
temperature.

\section{Qualitative analysis and numerical solutions of the DPE}

\subsection{General qualitative analysis}

For the normalized density $\rho (r)$, the DPE (3.8) reads (cf. \cite%
{Blinowski,Cahn}):
\begin{equation*}
\rho _{,rr}+\frac{2}{r}\rho _{,r}=\mu (\rho )-\mu _{\infty }.\eqno(4.1)
\end{equation*}%
Now, subscript means that the derivatives are taken with respect to the
normalized length variable $r$: in the whole section lengths and densities
are normalized with respect to $\displaystyle\mathcal{L}_{\alpha }=\sqrt{%
\frac{P_{c}}{\lambda \rho _{c}Z_{c}}}$ and critical density $\rho _{c}$. The
quantity $\mu $ is the normalized chemical potential and $\mu _{\infty }=\mu
(\rho _{l})$. The explicit normalized form of the free energy $W$ is given
below for some typical equations of state. $Z_{c}$ and $P_{c}$ are the
critical compressibility ratio and pressure. We will look for a solution of
the DPE verifying the boundary conditions considered in Section 3.1, i.e., $%
\rho _{,r}(0)=\rho _{,r}(\infty )=0$, that represents a physically possible
density profile in the phase transition region describing the equilibrium of
a bubble (droplet) with its liquid (vapor). The qualitative analysis is
based on the 'mechanical' interpretation of Eq. (4.1) \cite{van Kampen}.
This may be regarded as the equation of a 'particle' of mass $1$ moving in
the ($\mu _{\infty }$-dependent) potential:
\begin{equation*}
U(\rho ,\mu _{\infty })=-(W(\rho )-\mu _{\infty }\rho ),\eqno(4.2)
\end{equation*}%
with the 'viscous' time-dependent force $(2/r)\rho _{,r}$. In the following
we refer to the 'motion' of the particle as a solution of the Eq. (4.1)
starting from $r=0$, i.e., at the time zero, with zero velocity (cf. Section
3.1) and with a given initial density, i.e., a given initial position, $\rho
(0)$.

It is evident that a fundamental role in our problem is played by the
potential $U(\rho ,\mu _{\infty })$. For the sake of brevity, we limit
ourselves to some general considerations valid for any potential. We assume
under very general assumptions that:

\begin{itemize}
\item[(a)] for any temperature below the critical temperature there exists a
range of values for $\mu_{\infty}$, say $(\mu_{\infty}^{spi,bub},
\mu_{\infty}^{spi,drop})$, for which the potential (4.2) has three extrema $%
\rho_1(\mu_{\infty})<\rho_m(\mu_{\infty})<\rho_2(\mu_{\infty})$; $\rho_1$
and $\rho_2$ are the maxima and $\rho_m$ is a minimum;

\item[(b)] for $\mu _{\infty }=\mu _{\infty }^{spi,bub}$ (or $\mu _{\infty
}=\mu _{\infty }^{spi,drop}$), at the so-called \textit{spinoidal} limit
\cite{van Carey}, the potential has a maximum only in $\rho _{1}$ (or $\rho
_{2}$) and an inflexion point in $\rho _{2}$ (or $\rho _{1}$);

\item[(c)] the value of the chemical potential at the equilibrium between
phases with planar interface (saturation) $\mu_{\infty}^{sat}$ lies in the
interval $(\mu_{\infty}^{spi,bub},\mu_{\infty}^{spi,drop})$; for $%
\mu_{\infty}=\mu_{\infty}^{sat}$ the values of the potential is the same at
the two maxima (Maxwell rule), i.e., $U(\rho_1)=U(\rho_2)$;

\item[(d)] for $\mu _{\infty }^{spi,bub}<\mu _{\infty }<\mu _{\infty }^{sat}$
the value of the potential in $\rho _{2}$ is always lower than that in $\rho
_{1}$ , i.e., $U(\rho _{1})>U(\rho _{2})$ (bubble case); if $\mu _{\infty
}^{sat}<\mu _{\infty }<\mu _{\infty }^{spi,drop}$, the opposite happens: the
value of the potential in $\rho _{2}$ is higher than that in $\rho _{1}$,
i.e., $U(\rho _{1})<U(\rho _{2})$ (droplet case) \footnote{%
The simplest model is obtained by developing the free energy functional,
near the critical point, as a function of $\rho $; the result is $W(\rho
)=(\rho -a_{1})^{2}(\rho -a_{2})^{2}$ (with $a_{1}$ and $a_{2}$ constants
corresponding respectively to the homogeneous gas and liquid density), which
satisfies the properties (a)-(d) as it can be simply verified.}.
\end{itemize}

From assumptions (a)-(d) is clear that a 'motion' starting with initial
position $\rho(0)$ in the region just at the right (left) of the first
maximum $\rho_1$ (second maximum $\rho_2$) of the potential for the case of
the bubble (droplet) rolls down to the minimum $\rho_m$ and then goes up
towards the second maximum $\rho_2$ (first maximum $\rho_1$) where three
different cases are possible:

\begin{itemize}
\item[1)] The velocity is insufficient to reach the second maximum $\rho
_{2} $ (first maximum $\rho _{1}$); this happens because of the dissipative
term in Eq. (4.1). In this case the 'particle' reverses its velocity before
reaching the second maximum (first maximum) and falls towards the minimum
where it will be trapped with damped oscillations when $r$ tends to $\infty $%
;

\item[2)] The velocity is sufficient to reach the second maximum (first
maximum); then the 'particle' arrives in general with a finite velocity at
this point, then it falls down in the region of high density $\rho>\rho_2$
(low density $\rho<\rho_1$) after the maximum, its velocity grows and the
density reaches the co-volume (zero) limit in a finite time;

\item[3)] The velocity is exactly that one sufficient to reach the second
maximum (first maximum) (separatrix solution), so that the 'particle'
reaches the second maximum (first maximum) in an infinite time, i.e., for $r$
tending to $\infty$; at the maximum the velocity $\rho_{,r}$ is zero, so
that the boundary conditions stated in Section 3.1 are satisfied.
\end{itemize}

We will call $\rho _{f}$ the value for $\rho $ at which the potential
attains the lower maximum, i.e., $\rho _{f}=\rho _{1}$ if $\mu _{\infty
}^{sat}<\mu _{\infty }<\mu _{\infty }^{spi,drop}$, and $\rho _{f}=\rho _{2}$
if $\mu _{\infty }^{spi,bub}<\mu _{\infty }<\mu _{\infty }^{sat}$. Obviously
only the case 3) is physically significant. The force acting on the
'particle' is zero when $r$ tends to $\infty $ as the maximum is an extremum
of the potential. From Eq. (4.2) it follows:
\begin{equation*}
\mu (\rho _{f})=\mu _{\infty }=\mu (\rho _{\infty }).
\end{equation*}%
This implies that (we assume that locally for large [small] values of
density the function $\mu (\rho )$ is invertible), $\rho _{f}=\rho _{\infty
} $.

So the solution of the DPE is simply found if we examine the solution of the
equivalent mechanical problem when starting a 'motion' with an initial value
$\rho (0)$ \textit{such that we have exactly the separatrix solution}. So
for a fixed $\mu _{\infty }$ the only unknown quantity is $\rho (0)$ we will
find when we solve numerically Eq. (4.1).

We note that far from saturation the equality of (classical) chemical
potentials is not satisfied because in Eq. (4.1) we do not have zero on the
RHS when $r$ tends to zero. So for a small bubble an additional contribution
due to curvature appears in the Gibbs energy if we want to preserve the
classical relationship.

Near the saturation limit the problem becomes substantially different: in
the case of bubbles the 'particle' starts its motion in the neighborhood of $%
\rho _{1}$ and tends to reach $\rho _{f}=\rho _{2}$; its velocity has to be
relatively large and therefore dissipation could be very large and inhibit
the approach to the second maximum. However the system can react in a
different way as a full separatrix solution linking the two maxima exists
exactly at $\mu _{\infty }=\mu _{\infty }^{sat}$. In fact at saturation the
two maxima have the same height (Maxwell rule) and the separatrix starting
from the first maximum and ending over the second is the correct solution%
\footnote{%
This happens because the dissipation, i.e., the term $(2/r)\rho _{r}$,
cannot play any role because the escape time from the first maximum is now
infinite and this term tends to zero as $r$ tends to infinity.}. This is
exactly the Van Kampen solution for the planar interface \cite{van Kampen}.

\subsection{Free energy and $\protect\lambda $ value in some simple models}

The free energy functions can be chosen for the problem of phase transition
by modelling them on an appropriate equation of state for the single fluid.
The properties (a)-(d) will be shared by many approximate models of the long
range effects such as the van der Waals or the Berthelot equations of state.
In the following we study only the class of two-parameter equations of state
though the method is applicable also to more general equations. This class
includes several models \cite{Rocard} with the standard expression $a/V^{2}$
for the internal pressure and also some more sophisticated models involving
further correction of the internal pressure \cite{Peng}. However the form of
the free energy is substantially the same leading only to some relevant but
quantitative changes. We assume that the equation of state can be written in
a general form as:
\begin{equation*}
P(\rho ,T)=\rho \, tf(\beta \rho )-\alpha \rho ^{2}g(\beta \rho ),\eqno(4.3)
\end{equation*}%
where $t$ is the reduced temperature, i.e., $t=T/T_{c}$; $\displaystyle%
\alpha =\frac{a\rho _{c}}{\mathcal{R}_{M}T_{c}}$ (with $\mathcal{R}_{M}=%
\mathcal{R}/\mathcal{M}$, $\mathcal{R}$ the gas constant, $\mathcal{M}$ the
molecular weight) and $\beta =\rho _{c}\ b$ are the generalized normalized
van der Waals coefficients (which can depend on temperature $T$); $f$ and $g$
are suitable independent functions. In Table 1 we list the equations of
state studied in this paper, the form of functions $f$ and $g$ and the
values of the normalized coefficients $\alpha $ and $\beta $. The normalized
volume free energy, derived from the above equation of state, is:
\begin{equation*}
W(\rho )=\rho \, F_{0}(t)-\rho \, t\int_{\rho }^{\rho _{0}}f(\beta \rho )\,%
\frac{d\rho }{\rho }-\rho \, \alpha \int_{\rho }^{\rho _{0}}g(\beta \rho
)\,d\rho ,\eqno(4.4)
\end{equation*}%
where $\rho _{0}$ is a reference state at which free energy is $\rho
_{0}F_{0}(t)$. With regards to $\lambda $, Evans \cite{Evans} shows under
very general assumptions that:
\begin{equation*}
\lambda (\rho )=\frac{k_{B}T}{12}\int c[r;\rho ]\, r^{4}dr,\eqno(4.5)
\end{equation*}%
where $k_{B}$ is the Boltzmann constant and $c[r;\rho ]$ is the direct
correlation function of the fluid. This implies that $\lambda $ is a
functional of the field $\rho $. However under some degree of approximation
\cite{Ebner} an expression for the direct correlation function is given by
the Percus-Yevik equation (PY-equation) \cite{Percus}:
\begin{equation*}
c[r;\rho ]=g[r;\rho ](1-e^{-\frac{\mathcal{V}}{k_{B}T}}),
\end{equation*}%
where $g[r;\rho ]$ is the pair correlation function and ${\nu }$ is the
intermolecular potential. The expressions $c[r]$ and $g[r]$ can be
numerically found in this case \cite{Ebner}. In order to evaluate $\lambda $%
, we make the drastic approximation of taking $g[r]=\Theta (r-r_{0})$, where
$\Theta $ is the Heaviside function and $r_{0}$ is typically of the order of
magnitude of the molecular radius. \vskip0.8cm

\begin{center}
{\scriptsize
\begin{tabular}{|l|c|c|c|c|}
\hline
&  &  &  &  \\
State Equation & $\alpha$ & $\beta$ & $f$ & $g$ \\
&  &  &  &  \\ \hline\hline
&  &  &  &  \\
van der Waals & $\frac{9}{8}$ & $\frac{1}{3}$ & $\frac{1}{1-x}$ & $1$ \\
&  &  &  &  \\ \hline
&  &  &  &  \\
Rocard 1 & $1.254$ & $0.430$ & $\frac{1}{\left(1-\frac{x}{2}\right)^2}$ & $1$
\\
&  &  &  &  \\ \hline
&  &  &  &  \\
Rocard 2 & $1.319$ & $0.478$ & $\frac{1}{\left(1-\frac{x}{3}\right)^3}$ & $1$
\\
&  &  &  &  \\ \hline
&  &  &  &  \\
Rocard 3 & $1.500$ & $0.617$ & $e^{-x}$ & $1$ \\
&  &  &  &  \\ \hline
&  &  &  &  \\
Rocard 4 & $1.370$ & $0.515$ & $\frac{e^{-x}}{1-\frac{x^2}{8}}$ & $1$ \\
&  &  &  &  \\ \hline
&  &  &  &  \\
Berthelot & $\frac{9}{8}\frac{1}{t}$ & $\frac{1}{3}$ & $\frac{1}{1-x}$ & $1$
\\
&  &  &  &  \\ \hline
&  &  &  &  \\
Eberhart-Schnyders & $\frac{9}{8}\frac{1}{t^{\frac{1}{2}}}$ & $\frac{1}{3}$
& $\frac{1}{1-x}$ & $1$ \\
&  &  &  &  \\ \hline
&  &  &  &  \\
Peng-Robinson & $1.489(1-k(1-t^2))$ & $0.253$ & $\frac{1}{1-x}$ & $\frac{1}{2%
\sqrt{2}}\ln\frac{1-x-\sqrt{2}}{x-1-\sqrt{2}}$ \\
&  &  &  &  \\ \hline
\end{tabular}
}
\end{center}

\textbf{Table 1.} \ {\small Scaled parameters $\alpha$ and $\beta$ and
functions $f$ and $g$ for the equations of state considered in this paper }

Developing the Boltzmann factor in the PY-equation to the first order in ${%
\nu }$, this approximation leads to the standard results of Cahn and
Hilliard \cite{Cahn} or equivalently of Van Kampen \cite{van Kampen} (cf.
also Gouin \cite{Gouin2}) in which:
\begin{equation*}
\lambda \sim \int_{r_{0}}^{\infty }{\nu }(r)r^{4}dr,\eqno(4.6)
\end{equation*}%
We note that the normalizing length $\mathcal{L}_{\alpha }$, can be related
to $\lambda $ in the following way. In fact we can write $\mathcal{L}%
_{\alpha }=\sqrt{\alpha }\mathcal{L}$ where $\alpha =a\rho _{c}/RT_{c}$ and
(cf. \cite{Cahn}):
\begin{equation*}
\displaystyle\mathcal{L}^{2}=\displaystyle\frac{\displaystyle%
\int_{r_{0}}^{\infty }{\nu }(r)r^{4}dr}{\displaystyle\int_{r_{0}}^{\infty }{%
\nu }(r)r^{2}dr}.\eqno(4.7)
\end{equation*}%
We note that this last length depends strongly on the intermolecular
potential: for example for the $12-6$ Lennard-Jones potential, we have $%
\mathcal{L}=\sqrt{11/7}r_{0}$ in which $r_{0}\equiv r_{_{LJ}}$ is the
Lennard-Jones radius.

\subsection{Numerical results}

We integrate Eq. (4.1) by a Bulirsh-Stoer integrator \cite{Press} starting
from an arbitrary guess value of $\rho (0)$ with $\rho _{,r}(0)=0$. The
values of $\rho (0)$ are then adapted with a sequence of integrations until
a good approximation to the separatrix is reached. Near saturation, the
accuracy of the method is limited by the machine precision because the
starting density differs from the first maximum density $\rho _{1}$ by
progressively smaller quantities. We were limited to the usual double
precision ($10^{-16}$) of our computing device. We find that this precision
is sufficiently high to catch the main features of the approach to
saturation. To test the procedure we have verified that, starting from the
last point and going back toward small radii the 'particle' follows almost
exactly the same motion (backward in time). As a further test we have
verified that integral relation Eq. (3.11) is satisfied within the error
introduced by numerical integration.

\begin{figure}[h]
\begin{center}
\includegraphics[width=14cm]{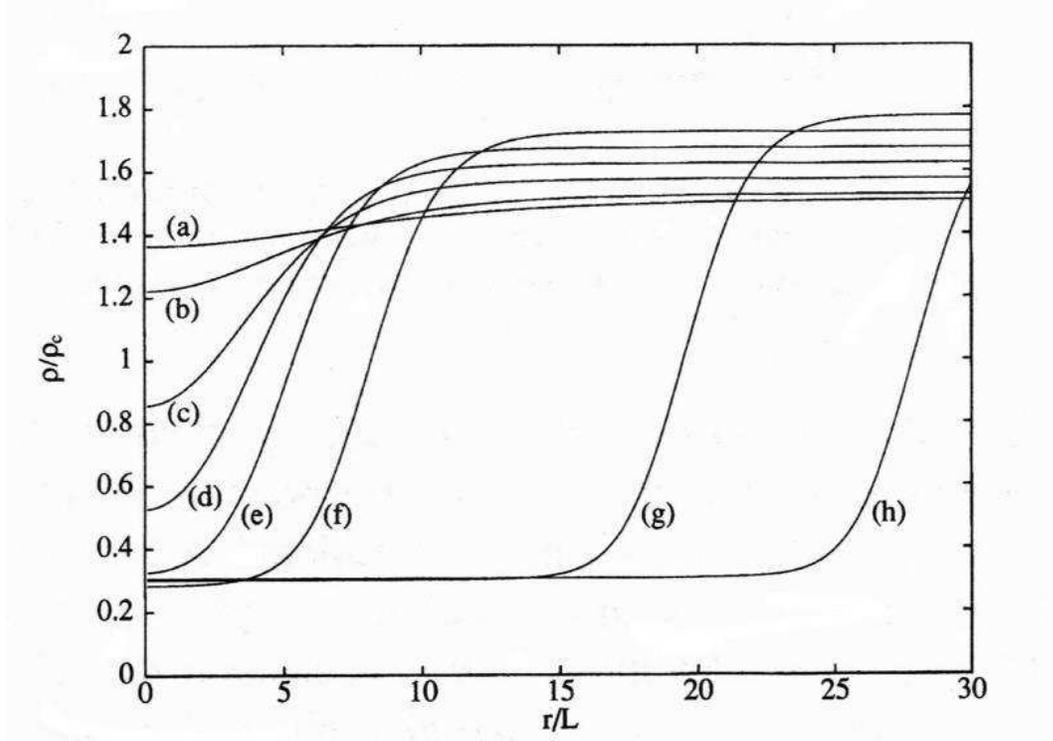}
\end{center}
\caption{Density profiles for van der Waals equations of state for different
values of $\protect\mu _{\infty }$ (or $\protect\rho _{\infty }$) at $t=0.85$
obtained by numerical solution of the DPE. Towards saturation, cf. the
curves labeled from (e) to (h), the bubble clearly develops a bulk vapor
phase in its interior. Near the minimal nucleation radius, cf. the curve
(d), the bulk vapor phase disappears. In the spinoidal limit, cf. the curves
from (c) to (a), the radius of the bubble again grows but no bulk vapor
phase exists and the density jump rapidly decreases until it disappears at
the so-called spinoidal composition in which only the liquid phase exists.}
\label{fig1}
\end{figure}
Typical density profiles are shown in Figure 1 for the van der Waals
equation of state. The two parts of the nucleation phenomenon, i.e., the
homogeneous nucleation and the growth of bubble approaching saturation, are
clearly seen in this figure. The density profile tends to be diffuse towards
the spinoidal limit so that the density jump $\rho _{\infty }-\rho (0)$ goes
to zero (cf. curves (a), (b) and (c) in Figure 1). Approaching saturation, a
region of almost-constant density appears before the phase transition
region, i.e. a bulk vapor phase, and the density jump is practically
constant (cf. curves (f), (g) and (h) in Figure 1).
\begin{figure}[h]
\begin{center}
\includegraphics[width=14cm]{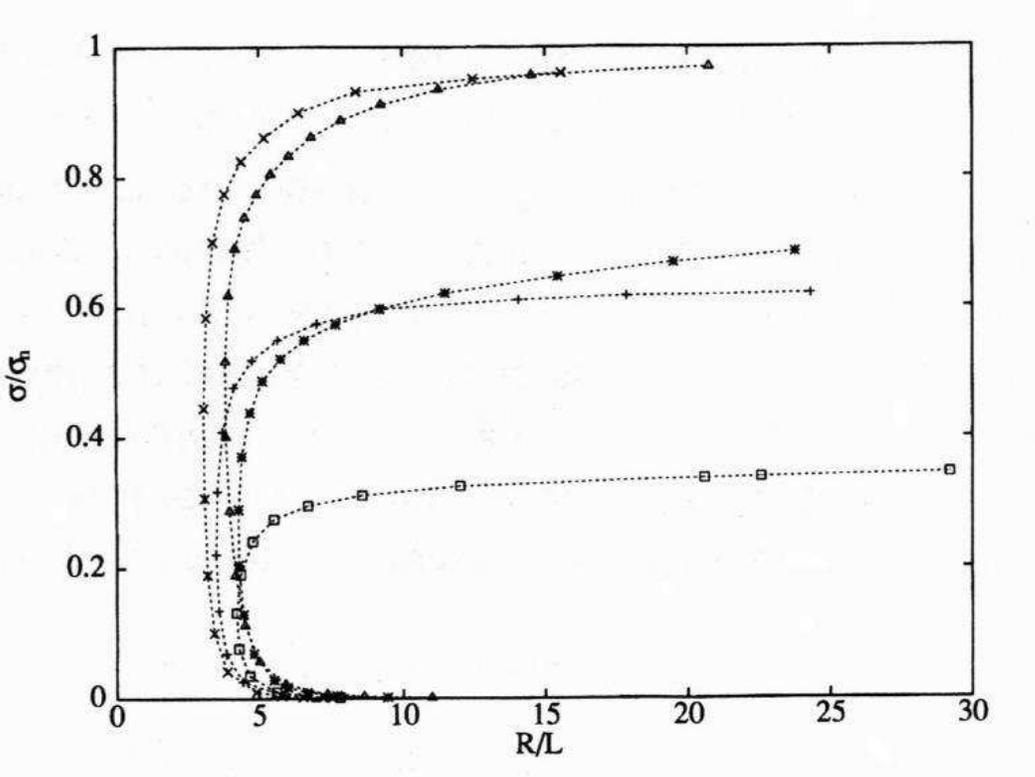}
\end{center}
\caption{Normalized surface tension $\protect\sigma /\protect\sigma _{n}$
where $\protect\sigma _{n}=\protect\rho _{c}R_{\mathcal{M}}T_{c}\mathcal{L}$
vs normalized equilibrium radius $R/\mathcal{L}$ of microscopic bubbles at $%
t=0.85$ for the following equations of state: van der Waals $\Box $; Rocard
4 $\star $; Berthelot $\times $; Peng-Robinson $\triangle $;
Eberhart-Schnyders $+$. The end-point of each curve toward the saturation
have to be considered equivalent in the sense of saturation proximity (cf.
the text).}
\label{fig2}
\end{figure}
In Figure 2 we show the dependence of $\sigma $ on $R$ (as defined in
Section 3.4) numerically evaluated at the reduced temperature $t=0.85$ for
some of the equations of state listed in Table 1. The results are
qualitatively similar: for small values of $\mu _{\infty }$ near the
spinoidal value $\mu _{\infty }^{bub,spi}$ (i.e. the lower part of the curve
in Figure 2), $\sigma $ initially vanishingly small, grows slowly whereas
the radius \textit{decreases} towards a minimal value that we identify as
the 'minimal nucleation radius' $R_{m}$. At $R_{m}$ the surface tension
suddenly increases and the density at the center of the bubble reaches the
range of the vapor phase (cf. also curves (d) and (e) in Figure 1). Above
the minimal nucleation radius, $\sigma $ approaches within a few minimal
radii the planar interface value and, as shown by Fisher and Israelachivli
experiments, remains substantially constant \cite{Fisher}. We note that
quantitatively, the van der Waals equation of state gives the lowest value
of planar interface surface tension, the Berthelot and the Peng-Robinson
give the highest values. The behavior for other values of temperature is
qualitatively the same. We note also that the curves end at different radii
as they approach saturation; these radii have to be considered as equivalent
because the 'distance' from saturation, measured by the difference between
the first maximum of the potential and the density at the center of the
bubble, is in all cases less than $10^{-16}$. It could be implied by this
circumstance that different equations of state reach saturation at different
radii. Moreover we note that curves corresponding to radii between the
spinoidal and the minimal critical radius values can be considered as
lacking physical meaning because, no physical bubble can exist in this
regime, except fluctuations of typical radius $R$ (cf. \cite{Cahn}). So the
'minimal nucleation radius' $R_{m}$ marks the last values, before the
spinoidal, to which one can give physical meaning.\newline
In Figure 3 we plot the (equivalent) interface thickness defined in Eq.
(3.20) as a function of $R$ for some of the different equations of state,
always at reduced temperature $t=0.85$. At the start of nucleation
(spinoidal limit), $\delta $ is very large (diffuse interface) and of the
same order of magnitude as $R$: density fluctuations have a unique length
scale. When $R$ has reached the minimal nucleation radius, $\delta $ quickly
tends to a practically constant value with very small deviation from the
limit value at saturation. Again all the studied equations of state have the
same qualitative behavior. Moreover, we note that the curves between the
spinoidal limit and the minimum radius are essentially the same for all the
equations of state used here (cf. Figure 3). \newline
We note that no qualitative differences appear when the temperature is
changed. It is also interesting to note that the ratio $(\delta /R)_{m}$ is,
within the numerical accuracy, a constant independent of $t$: its value is
reported in Table 2 for all equations of state. Moreover this constant is
practically the same for all the equations of state studied here, i.e., $%
(\delta /R)_{m}\sim \;0.42-0.44$. The justification of this phenomenon can
be that, the shell-like quantities as surface tension, radius, etc., being
defined through an integral over the thickness of the interface, the
differences between the equations of state will be smeared by integration,
giving similar results.

\begin{center}
{\scriptsize
\begin{tabular}{|l|c|c|c|}
\hline
&  &  &  \\
State Equation & $(\delta/R)_m$ & $\delta_{\infty}/\delta_m$ & $%
\rho_{\sigma}^{\infty}/\rho_{\sigma}^{m}$ \\
&  &  &  \\ \hline
&  &  &  \\
van der Waals & $0.427$ & $0.771$ & $0.716$ \\
&  &  &  \\ \hline
&  &  &  \\
Rocard 1 & $0.443$ & $0.778$ & $0.712$ \\
&  &  &  \\ \hline
&  &  &  \\
Rocard 2 & $0.435$ & $0.785$ & $0.739$ \\
&  &  &  \\ \hline
&  &  &  \\
Rocard 3 & $0.443$ & $0.768$ & $0.716$ \\
&  &  &  \\ \hline
&  &  &  \\
Rocard 4 & $0.432$ & $0.795$ & $0.773$ \\
&  &  &  \\ \hline
&  &  &  \\
Berthelot & $0.426$ & $0.768$ & $0.710$ \\
&  &  &  \\ \hline
&  &  &  \\
Eberhart-Schnyders & $0.428$ & $0.761$ & $0.683$ \\
&  &  &  \\ \hline
&  &  &  \\
Peng-Robinson & $0.439$ & $0.750$ & $0.767$ \\
&  &  &  \\ \hline
\end{tabular}%
}
\end{center}

\textbf{Table 2. }\ {\small Numerically evaluated ratios $(\delta/R)_m$, $%
\delta_{\infty}/\delta_m$ and $\rho_{\sigma}^{\infty}/\rho_{\sigma}^m$ for
the equations of state reported in Table 1. The ratios are substantially
independent of the temperature in the limit of the algorithm error
introduced in the numerical solutions of the DPE equation.}

\begin{figure}[h]
\begin{center}
\includegraphics[width=14cm]{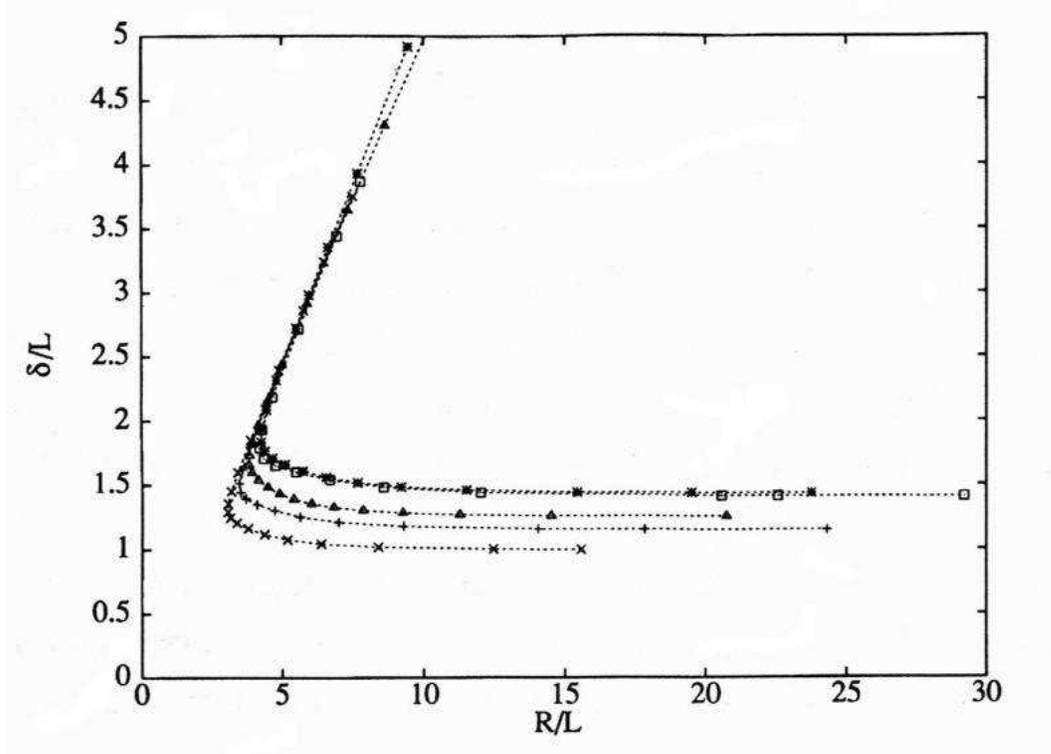}
\end{center}
\caption{Normalized thickness $\protect\delta/\mathcal{L}$ of the interface
vs normalized equilibrium radius $R/\mathcal{L}$ of microscopic bubbles at $%
t=0.85$ for the following equations of state: van der Waals $\Box$; Rocard 4
$\star$;Berthelot $\times$; Peng-Robinson $\triangle$; Eberhart-Schnyders $+$%
.}
\label{fig3}
\end{figure}

\begin{figure}[h]
\begin{center}
\includegraphics[width=14cm]{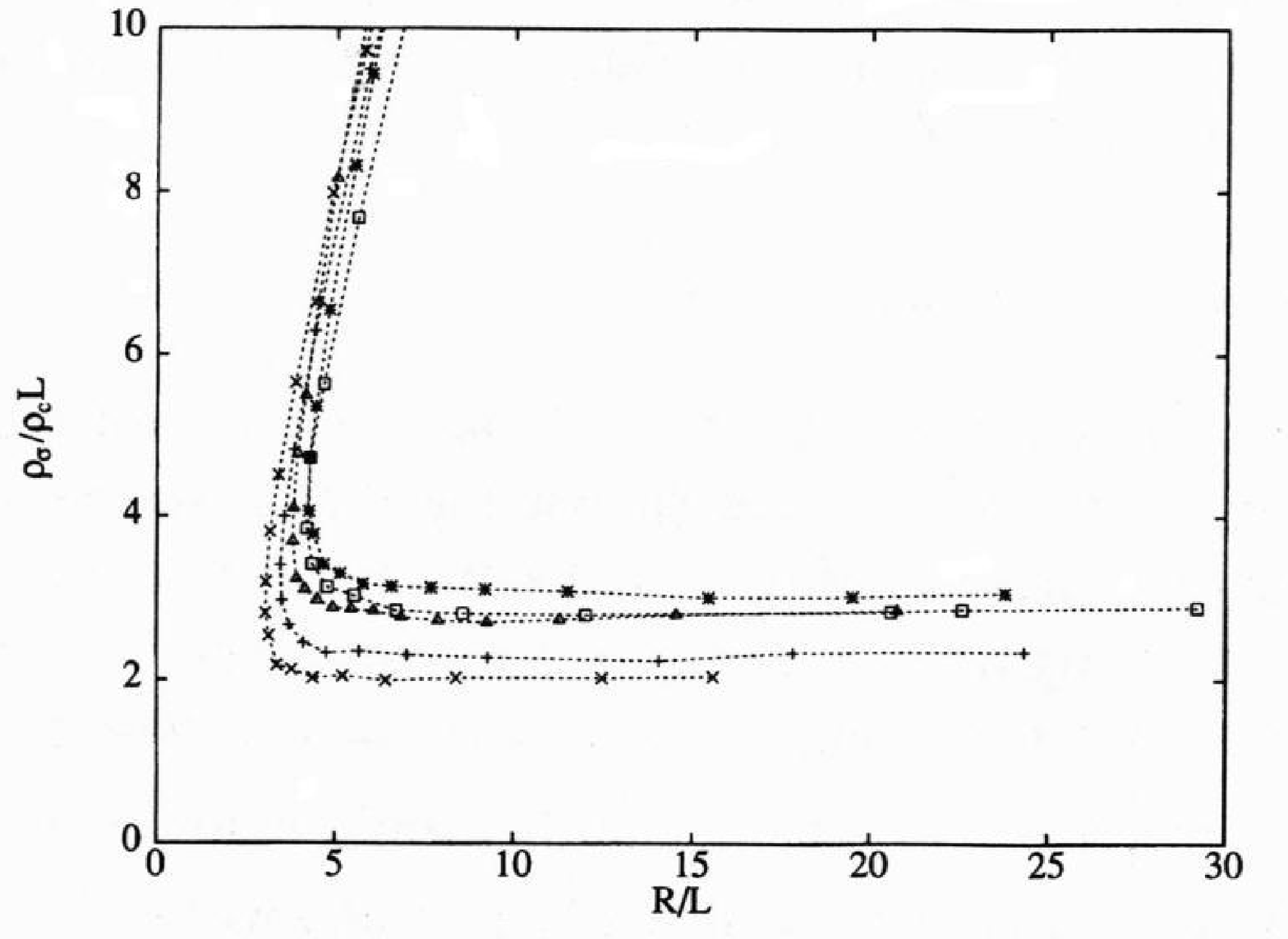}
\end{center}
\caption{Normalized surface mass density $\protect\rho_{\protect\sigma}/(%
\protect\rho_c\mathcal{L})$ vs normalized equilibrium radius $R/\mathcal{L}$
of microscopic bubbles at $t=0.85$ for the following equations of state: van
der Waals $\Box$; Rocard 4 $\star$; Berthelot $\times$; Peng-Robinson $%
\triangle$; Eberhart-Schnyders $+$.}
\label{fig4}
\end{figure}

In Figure 4 we plot the surface mass density versus $R$ at $t=0.85$ obtained
by using Eq. (3.22) for the same equations of state than in Figures 2-3. We
see that surface mass is an increasing function of $R$ going towards the
spinoidal limit. This behavior is obviously due to the increase of $\delta $
in the same limit. Again we note that no qualitative differences appear when
the temperature is changed. Moreover the ratios $\delta _{\infty }/\delta
_{m}$ and $\rho _{\sigma }^{\infty }/\rho _{\sigma }^{m}$ are independent of
$t$. Their values, listed in Table 2 for all the equations of state, differ
by less than $15\%$.

\section{Minimal nucleation radii: theoretical predictions and comparison
with experiments}

The comparison of the above results with the experimental values reported in
literature is a rather difficult task because information for many data can
be only indirectly recovered. Moreover, as shown in Section 4.2, Eqs
(4.5-4.7), all the predictions are critically dependent on the parameters of
the intermolecular potential; in our case a fundamental role is played by $%
\mathcal{L}$ that sets the length scales. Again the correct prediction of
surface tension is also not simple in the sophisticated microscopic
statistical theories (cf. the data reported in \cite{Evans}) because,
especially at very small scales, the details of the intermolecular
potential, and so the particular molecular micro-structure, are relevant in
the determination of the exact values of surface tension. We prove in this
paper that rather good predictions of surface tension are also possible in
the framework of the second gradient theory.

Moreover we are able to produce predictions of nucleation radius. In
particular, by evaluating $R_{m}$, we establish a lower bound to the
experimental nucleation radii reported in the literature. We remark that
very few data are found in the literature concerning bubbles (cf. \cite{van
Carey}), so that we are obliged to compare these predictions with data for
the droplets found in Lamer and Pound or Kumar \cite{Kumar,Lamer}.%
\newline
In order to choose, for every substance, the most suitable equations of
state to predict the minimal nucleation radius, we use the planar interface
surface tension values in the following way: by following Cahn and Hilliard,
we fix $\mathcal{L}$ (cf. the next subsection) while among the considered
equations of state, we choose that which predicts $\sigma _{\infty }$ values
sufficiently accurately (for many substances this estimate holds within $%
90\% $ of the experimental value). Finally we can predict the thickness of
the interface and the surface mass density. We note that these data are
relatively scarce especially the latter \cite{Alts,Isola3}.

\subsection{The planar interface surface tension.}

\begin{center}
{\scriptsize
\begin{tabular}{|l|c|c|c|c|}
\hline
&  &  &  &  \\
Substance & $t_{exp}$ & $\sigma^{exp}_{\infty}\;\;\;[\frac{dyn}{cm}]$ & $%
\sigma^{vdw}_{\infty}\;\;\;[\frac{dyn}{cm}]$ & $\alpha_{exp}$ \\
&  &  &  &  \\ \hline\hline
&  &  &  &  \\
Ar & $\sim 0.5$ & $13.1$ & $23.9$ & $3.24$ \\
&  &  &  &  \\ \hline
&  &  &  &  \\
CO$_2$ & $0.95$ & $1.16$ & $0.93$ & $1.51$ \\
&  &  &  &  \\ \hline
&  &  &  &  \\
N$_2$ & $\sim 0.6$ & $9.80$ & $8.25$ & $1.50$ \\
&  &  &  &  \\ \hline
&  &  &  &  \\
O$_2$ & $\sim 0.6$ & $13.2$ & $11.4$ & $1.46$ \\
&  &  &  &  \\ \hline
&  &  &  &  \\
H$_2$ & $\sim 0.6$ & $2.09$ & $2.29$ & $1.34$ \\
&  &  &  &  \\ \hline\hline
&  &  &  &  \\
H$_2$O & $0.5$ & $68.5$ & $63.8$ & $1.83$ \\
&  &  &  &  \\ \hline
&  &  &  &  \\
NH$_3^*$ & $\sim 0.7$ & $23.4$ & $26.3$ & $1.73$ \\
&  &  &  &  \\ \hline
&  &  &  &  \\
CH$_3$OH & $\sim 0.6$ & $20.1$ & $20.1$ & $1.91$ \\
&  &  &  &  \\ \hline
&  &  &  &  \\
C$_2$H$_5$OH$^*$ & $\sim 0.5$ & $22.7$ & $24.7$ & $1.70$ \\
&  &  &  &  \\ \hline
&  &  &  &  \\
C$_2$H$_6^*$ & $\sim 0.6$ & $16.2$ & $15.3$ & $1.50$ \\
&  &  &  &  \\ \hline
&  &  &  &  \\
C$_6$H$_{12}$ & $0.5$ & $25.5$ & $16.5$ & $1.63$ \\
&  &  &  &  \\ \hline
&  &  &  &  \\
CCl$_4$ & $\sim 0.5$ & $29.8$ & $18.6$ & $1.64$ \\
&  &  &  &  \\ \hline
&  &  &  &  \\
CF$_4$ & $\sim 0.9$ & $5.0$ & $0.87$ & $-$ \\
&  &  &  &  \\ \hline
\end{tabular}
}
\end{center}

\textbf{Table 3.}\ {\small Planar interface surface tension for some
substances evaluated in the case of van der Waals equation of state and the
corresponding experimental value. In the last column of the Table is
reported the scaled parameter $\displaystyle\alpha _{exp}=\frac{a_{exp}\rho
_{c}}{\mathcal{M}RT_{c}}$. For experimental planar interface surface tension
cf. \cite{Gaz,CRC}. The upper part of the Table shows substances for which
Lennard-Jones radius $r$}$_{{\small _{LJ}}}${\small \ is known with good
approximation and so it is used to calculate $\sigma _{\infty }$ by
evaluating $\mathcal{L}$ from $\displaystyle(11/7)^{\frac{1}{2}}r$}$_{%
{\small _{LJ}}}${\small . In the lower part of the Table (down the double
line) the covolume radius $r_{co}$ is used to evaluate $\mathcal{L}$ and
then $\sigma _{\infty }$. In the cases marked by $^{\ast }$ a better fitting
was made by taking $\mathcal{L}=1.25\;r_{co}$.}

In Table 3 we show collected experimental data for different substances.
Values for the planar interface surface tension are shown together with the
values evaluated by second gradient theory in the case of the van der Waals
equation of state. The theoretical predictions were made by using in some
cases the Lennard-Jones relation between $r_{0}\equiv r_{_{LJ}}$ and $%
\mathcal{L}$ while in others we have simply put $\mathcal{L}=r_{0}=r_{co}$,
where $r_{co}$ is the covolume radius (we note that in general $%
r_{_{LJ}}>r_{co}$ but their ratio is lower than $(11/7)^{\frac{1}{2}}\sim
1.25$).

The prediction of the second gradient theory is generally within $20\%$ of
the experimental value: this can be considered as a good result, comparable
with the non-local model approach (cf. \cite{Davis} and \cite{Evans}). In
many cases our predictions are very accurate. We note however that are
remarkable exceptions: 1) the Ar value is larger by about a factor two with
respect to the experimental value; the same holds true in respect of the
surface tension for other noble gases; 2) the values for C$_{6}$H$_{12}$ and
CCl$_{4}$ are smaller almost by a factor of two (about $60\%$ of
experimental values), the same is true in minor form also for other carbon
compounds such as C$_{4}$H$_{10}$ and C$_{3}$H$_{8}$ and most dramatically
for CF$_{4}$, where the ratio is about $1/6$.

The case of Argon and the other noble gases can be understood if we look to
the last column of Table 3 in which we report the scaled experimental value
of $\displaystyle\alpha _{exp}=\frac{a_{exp}\rho _{c}}{\mathcal{M}RT_{c}}$
representing the normalized value of the van der Waals constant $a$. For all
substances $\alpha _{exp}$ is between $1$ and $2$ whereas for Ar it is
larger than $2$ (about $3.2$); the same is true for other noble gases. This
large value implies that the values of surface tension evaluated with the
'van der Waals theory' are generally far from the experimental values. This
is a known result for Ar \cite{Evans}: the second gradient theory is
insufficient to reproduce the experimental value for the surface tension, as
higher-order gradients are necessary in the constitutive law for free energy
to give a correct prediction\footnote{%
We note that values of $\alpha _{exp}$ higher than for noble gases are
obtained for water and methyl alcohol, this is reflected, (cf.   Table 3,
caption), in the use of $r_{co}$ rather than $r_{_{LJ}}$ in the reported
theoretical predictions.} (cf. \cite{Ebner}).

As for point 2) we could increase the values of the surface tension, by
multiplying $r_{co}$ by $1.25$, i.e., the Lennard-Jones factor. This trick
works effectively for ethyl alcohol, ethane and ammonia, but for the other
substances the theoretical surface tension remains too small when compared
with the experimental value (we note also the case of CO$_2$ in which the
theoretical value is $79\%$ of the experimental value). So we conjecture
that other equations of state, that generally have a larger value of $%
\sigma_{\infty}$ (cf. Fig. 2), would describe carbon compounds better than
the van der Waals equation. For the sake of brevity we do not present the
theoretical data; we limit ourselves to saying that, by using a different
equation of state, the values of $\sigma_{\infty}$ can be predicted with the
same precision as shown in Table 3 for substances that are well described by
the van der Waals equation.

\subsection{The minimal nucleation radius and the other surface quantities}

\begin{figure}[h]
\begin{center}
\includegraphics[width=14cm]{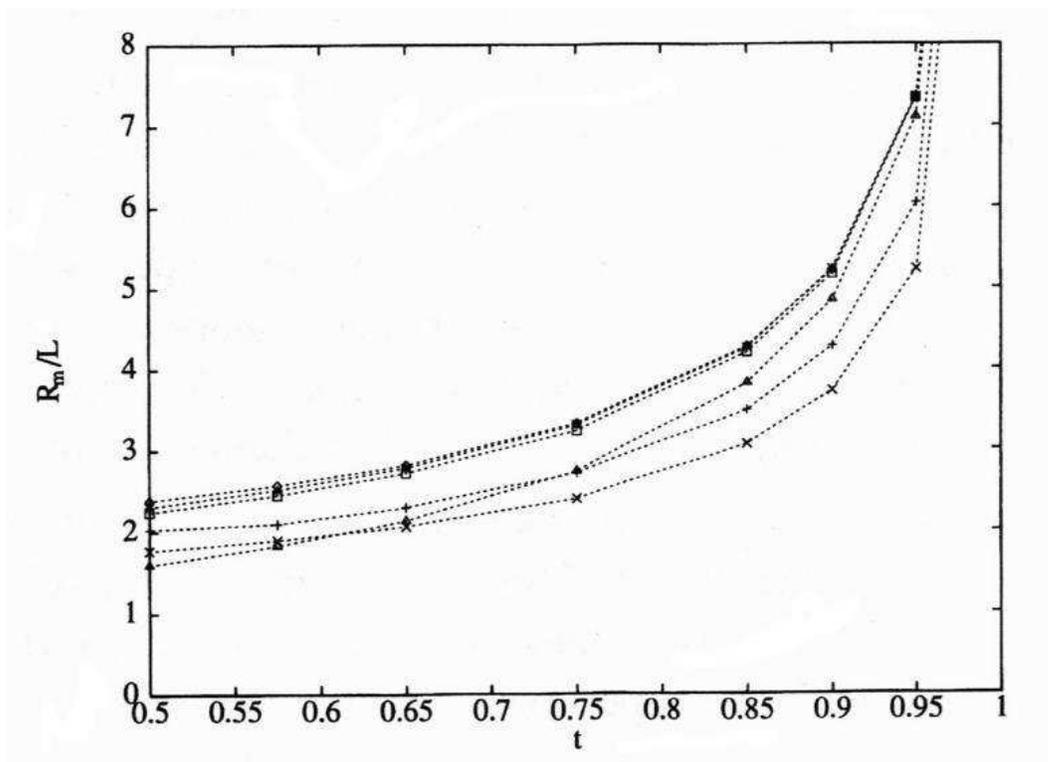}
\end{center}
\caption{Minimal nucleation radii vs reduced temperature in units of $%
\mathcal{L}$ in the interval $[0.5,1.0[$ for the following equations of
state: van der Waals $\Box$; Rocard 3 $\Diamond$; Rocard 4 $\star$;
Berthelot $\times$; Peng-Robinson $\triangle$; Eberhart-Schnyders $+$. State
equations Rocard 1 and 2 give minimal nucleation radii between Rocard 3 and
van der Waals cases.}
\label{fig5}
\end{figure}

In Figure 5 we display the normalized minimal nucleation radius in units of $%
\mathcal{L}$ for the equations of state in Table 1 versus reduced
temperature in the interval $[0.5,1.0[$. The interval chosen is the most
relevant for many substances because most of the experimental data on
surface tension exist in this range of reduced temperature (cf. \cite%
{Gaz,CRC}). At the lower temperatures the minimal radius tends to be a
constant of approximate value $\mathcal{L}$; the exact values could be
retrieved in principle by the method described in Section 4, but the strong
stiffness of the problem, induced by the rapid decrease to very small values
of the gas phase density, limits the precision of the method.

\begin{center}
{\scriptsize
\begin{tabular}{|l|c|c|c|}
\hline
&  &  &  \\
Substance & $R_m$ [\AA ] & $\delta_m$ [\AA ] & $\rho_{\sigma}^{m}\;10^{-7}%
\cdot [g/cm^2]$ \\
&  &  &  \\ \hline
&  &  &  \\
CO$_2$ & $22.3$ & $9.52$ & $9.4$ \\
&  &  &  \\ \hline
&  &  &  \\
N$_2$ & $19.7$ & $8.47$ & $5.7$ \\
&  &  &  \\ \hline
&  &  &  \\
O$_2$ & $18.4$ & $7.86$ & $7.4$ \\
&  &  &  \\ \hline
&  &  &  \\
H$_2$ & $15.6$ & $6.67$ & $0.4$ \\
&  &  &  \\ \hline\hline
&  &  &  \\
H$_2$O & $15.5$ & $6.63$ & $4.6$ \\
&  &  &  \\ \hline
&  &  &  \\
NH$_3^*$ & $20.8$ & $8.87$ & $4.5$ \\
&  &  &  \\ \hline
&  &  &  \\
CH$_3$OH & $20.2$ & $8.62$ & $5.0$ \\
&  &  &  \\ \hline
&  &  &  \\
C$_2$H$_5$OH$^*$ & $27.3$ & $11.6$ & $6.9$ \\
&  &  &  \\ \hline
&  &  &  \\
C$_2$H$_6^*$ & $24.9$ & $10.6$ & $4.7$ \\
&  &  &  \\ \hline
\end{tabular}%
}
\end{center}

\textbf{Table 4.}\ {\small Minimal nucleation radii for bubbles of some
substances evaluated at $t=0.85$ in the case of van der Waals equation of
state. In the other two columns are listed the predicted values of thickness
and surface mass density at minimal nucleation radius at the same reduced
temperature. The meaning of the division of the Table in upper and lower
part and the $^*$ near some substance is the same than in Table 3.}

We see that the Berthelot equation of state and the modification proposed by
Eberhart-Schnyders \cite{van Carey} predict radii rather smaller than other
equations of state. The Peng-Robinson equation of state gives radii near to
the van der Waals model at high values of $t$ and near the Berthelot model
values at about $t=0.5$. The family of all equations of state proposed by
Rocard give radii very near to the van der Waals value (for this reason only
two of these data are reported in Figure 5, in order to leave the figure
sufficiently clear). Obviously, when the critical temperature is approached
the minimal radius and the thickness tend to $\infty $.

\begin{center}
{\scriptsize
\begin{tabular}{|l|c|c|c|c|}
\hline
&  &  &  &  \\
Substance & State Equation & $R_m$ [\AA ] & $\delta_m$ [\AA ] & $%
\rho_{\sigma}^{m}\;10^{-7}\cdot [g/cm^2]$ \\
&  &  &  &  \\ \hline
&  &  &  &  \\
CH$_4$ & Eberhart-Schnyders & $13.1$ & $5.70$ & $2.0$ \\
&  &  &  &  \\ \hline
&  &  &  &  \\
C$_3$H$_8$ & Rocard 4 & $22.1$ & $9.54$ & $4.6$ \\
&  &  &  &  \\ \hline
&  &  &  &  \\
C$_3$H$_8$ & Eberhart-Schnyders & $18.1$ & $7.86$ & $3.8$ \\
&  &  &  &  \\ \hline
&  &  &  &  \\
C$_4$H$_{10}$ & Rocard 4 & $25.1$ & $10.8$ & $5.4$ \\
&  &  &  &  \\ \hline
&  &  &  &  \\
C$_4$H$_{10}$ & Eberhart-Schnyders & $20.5$ & $8.90$ & $4.5$ \\
&  &  &  &  \\ \hline
&  &  &  &  \\
C$_6$H$_{12}$ & Rocard 4 & $26.3$ & $11.3$ & $6.8$ \\
&  &  &  &  \\ \hline
&  &  &  &  \\
C$_6$H$_{12}$ & Peng-Robinson & $23.6$ & $10.2$ & $6.2$ \\
&  &  &  &  \\ \hline
&  &  &  &  \\
CCl$_4$ & Rocard 4 & $26.1$ & $11.2$ & $13.8$ \\
&  &  &  &  \\ \hline
&  &  &  &  \\
CCl$_4$ & Peng-Robinson & $23.4$ & $10.1$ & $12.6$ \\
&  &  &  &  \\ \hline
\end{tabular}%
}
\end{center}

\textbf{Table 5.} \ {\small Minimal nucleation radii for bubbles of some
substances (carbon compounds) evaluated at $t=0.85$ in the case of some
other equations of state. These will be chosen to reproduce the planar
interface surface tension experimental values at least to $80\%$. In the
other two columns are listed the predictions for thickness and surface mass
density at minimal nucleation radius at the same reduced temperature.}

In Table 4, the minimal nucleation radii at $t=0.85$ are listed for those
substances whose planar interface surface tension is well described by the
van der Waals equation of state. In the same table, we have added data on
the thickness of the interface and the surface mass density for the
substances cited therein. In Table 5, further theoretical predictions are
given, always at $t=0.85$, for substances whose planar interface surface
tension is well described by other equations of state. Thus we show that
radii, thickness and surface mass densities can be calculated very easily by
using the results found in \cite{Isola1}. We note that the minimal radius
ranges typically from $10$ to $30$ \AA ~ so that it is just a few molecular
radii. For example the molecules of C$_{6}$H$_{12}$ have a radius of about $%
5 $ \AA ; at $t=0.85$ the ratio between this value and predicted minimal
radius is about $4.7$, while at $t=0.5$ the minimal nucleation radius is $%
\sim 9.8$ \AA ~ and the ratio is about $2$.

Examples of predictions of the equilibrium radius as a function of $R$ (or $%
\mu _{\infty }$) by means of classical theory are given by van Carey \cite%
{van Carey}. However he limits himself to the case of 'large bubbles', the
only ones for which classical theory is valid. The experiment of Fisher and
Israelachvili \cite{Fisher} needs to be reinterpreted because the Laplace
formula, on which is based the Kelvin relation used to extrapolate the
radius, is not valid near the minimal nucleation radius \cite{Isola4}.
Moreover, as noted above, their results can be compared with a process of
homogeneous nucleation only after a thorough analysis of the effect of their
experimental apparatus. We note that as cited above C$_{6}$H$_{12}$ has a
predicted minimal nucleation radius of about $\sim 9.8$ \AA ~at the
temperature in which the experiment was performed ($t\sim 0.5$). It is
larger than the radii reported by Fisher and Israelachvili (we also note
that, on changing the equations of state or the evaluation of $\mathcal{L}$,
the minimal radius \textit{always} remains greater than their data). On the
other hand, data on homogeneous nucleation for very small radii exist for
droplets (cf. \cite{Kumar,Lamer}) and comparison with our data shows that
the nucleation radii of bubbles we predict are in some cases very close to
nucleation radius for the droplets; for example, in the case of water, the
Lamer and Pound data are in the range $8.0-8.9$ \AA ~at $t\sim 0.5$, and our
prediction is $R_{m}\simeq 8.3$ \AA ; the data for equilibrium radii
reported by Kumar for the CCl$_{4}$ at $t\simeq 0.5$ range between $12-17$
\AA , these are very close to those we calculated which range (at $t=0.5$)
between $12-14$ \AA , though this range of variation is due to the used
equation of state rather than to the supersaturation ratio.

Moreover we predict that the thickness of the interface is formed by few
molecular layers \cite{Evans} in the range from minimal nucleation radius to
saturation. A new insight of the physical problem is obtained when observing
that at small radii the interface constitutes a \textit{large} part of the
microscopic bubbles and cannot be confined in any Gibbs-Tolman idealized
zero thickness layer. Typically the numerically evaluated ratio between the
equivalent thickness at the minimal radius and the minimal radius itself is
of the order of $0.4$ for all the equations of state studied here. Finally
we note that the surface mass density values (see Tables 4 and 5) are of the
order of magnitude predicted by Alts and Hutter in their theory on water
\cite{Alts}.

\section{Conclusion}

The first three Sections of this paper give a complete account of the
results available in the literature concerning the theoretical treatment of
the phenomenon of homogeneous nucleation and the dependence of surface
tension on curvature.

However recently (cf. \cite{Isola4}) a rational definition for surface
tension and radius of curvature for spherical interfaces has been proposed
in terms of a solution of the density field, using a shell-like approach to
the definition of interface quantities. Moreover, thanks to the second
gradient theory, a deeper understanding of the relationship between the
Laplace formula and the capillary constant $\lambda $ has become possible.
In particular our investigations stem from a theoretical treatment in which:
\newline
1) an expression for surface tension is used which accounts for curvature
effects better than those found in the literature; \newline
2) the mechanical pressure is distinguished from the thermodynamic pressure.

In this way we obtained, using numerical methods, predictions for: \newline
1) the dependence of surface tension, equilibrium radius, thickness of the
spherical interface and surface mass density on chemical potential; \newline
2) the values of the minimal nucleation radius, i.e., the minimum
equilibrium radius possible for a bubble. Being founded either on
statistical mechanics or on second gradient theories, our predictions have a
clear meaning in a range of temperature close to the critical point.

Possible generalizations include the study of: \newline
(i) models in which $\lambda $ depends on $\rho $ locally or globally;
\newline
(ii) models in which higher density gradients appear, or a fully non-local
approach via integral equations; \newline
(iii) models suitable to describe the dynamics of phase transitions.

\textbf{Acknowledgements:}

We wish to thank A.\ Di Carlo, P.\ Seppecher and W.\ Kosi\'{n}ski for their
friendly criticism.\ This work was financially supported by the Universit%
\'{e} de Toulon et du Var, the Universit\'{e} of Aix-Marseille and the
Italian CNR.

\end{document}